# From colossal to zero: Controlling the Anomalous Hall Effect in Magnetic Heusler Compounds via Berry Curvature Design

Kaustuv Manna[1,*], Lukas Muechler[1,2,†], Ting-Hui Kao[1,3], Rolf Stinshoff[1], Yang Zhang[1], Johannes Gooth[1], Nitesh Kumar[1], Guido Kreiner[1], Klaus Koepernik[4], Roberto Car[2], Jürgen Kübler[1,5], Gerhard H. Fecher[1], Chandra Shekhar[1], Yan Sun[1], Claudia Felser[1,*]

[1]*Max Planck Institute for Chemical Physics of Solids, 01187 Dresden, Germany.*

[2]*Department of Chemistry, Princeton University, Princeton, New Jersey 08544, USA.*

[3]*Department of Physics, National Sun Yat-Sen University, Kaohsiung 804, Taiwan.*

[4]*IFW Dresden, P.O. Box 270116, 01171 Dresden, Germany.*

[5]*Institut für Festkörperphysik, Technische Universität Darmstadt, D-64289 Darmstadt, Germany.*

Since the discovery of the anomalous Hall effect (AHE), the anomalous Hall conductivity (AHC) has been thought to be zero when there is no net magnetization. However, the recently found relation between the intrinsic AHE and the Berry curvature predicts other possibilities, such as a large AHC in non-colinear antiferromagnets with no net magnetization but net Berry curvature. Vice versa, the AHE in principle could be tuned to zero, irrespective of a finite magnetization. Here, we experimentally investigate this possibility and demonstrate that, the symmetry elements of Heusler magnets can be changed such that the Berry curvature and all the associated properties are switched while leaving the magnetization unaffected. This enables us to tune the AHC from 0 $\Omega^{-1}$cm$^{-1}$ up to 1600 $\Omega^{-1}$cm$^{-1}$ with an exceptionally high anomalous Hall angle up to 12%, while keeping the magnetization same. Our study shows that the AHC can be controlled by selectively changing the Berry curvature distribution, independent of the magnetization.

In conventional metals or semiconductors, a transverse voltage is generated due to the Lorentz force when a magnetic field is applied perpendicular to an applied electric current. This effect leads to an anti-symmetric contribution to the off-diagonal electrical resistivity, which is generally known as the Hall effect. In ferro- or ferrimagnets it is believed that the spontaneous magnetization generates an additional transverse voltage due to the material's spin-orbit coupling (SOC), which causes carriers to be deflected by the magnetic moments of the host solid. This effect is known as the anomalous Hall effect (AHE) [1-3]. The AHE has two different contributions, an extrinsic contribution from scattering and an intrinsic contribution from the band structure [1,4]. For a long time it was believed that the anomalous Hall conductivity (AHC) scales with the sample's magnetization. According to this notion, any ferromagnetic material exhibits an AHE, but it is *zero* for an antiferromagnet, owing to the compensation of the magnetic sublattices ($M = 0$). Therefore, the AHE has been considered as the key signature of finite magnetization in ferromagnets or ferrimagnets. However, only recently, it was realized that the *intrinsic* contribution to the AHE is not directly related to the sample magnetization of a material but derives more generally from its net Berry curvature.

The Berry curvature distribution in materials is a property of the band structure that determines the topological aspects of a material [2,5-7]. A necessary condition for a finite net Berry curvature and thus a nonzero AHE is the absence of symmetries that reverse the sign of the local Berry curvature in the Brillouin zone (BZ) when reversing the sign of the momentum vector, e.g. time-reversal symmetry and mirror operations. In contrast to previous believes it is thus possible to control the Berry curvature and intrinsic AHE by suitable manipulations of symmetries and band structures, independent of the finite value of the magnetization [8]. As a result of these considerations, a strong AHC was predicted in the non-collinear antiferromagnetic systems like $Mn_3Ir$ [9], $Mn_3Ge$ and $Mn_3Sn$ [10] *etc*. and then experimentally observed in $Mn_3Ge$ [11] as well as Mn$_3$Sn [12,13]. Moreover, very recently, a large intrinsic AHE has been found in magnetic Weyl semimetals with broken time-reversal symmetry that depends on the separation of the Weyl nodes in momentum space [1,14-17]. In Weyl semimetals, the Weyl point acts as the monopole of Berry curvature, where a topological invariant, the Chern number, can be assigned to each Weyl node [18,19]. Therefore, one can tune the AHC via the symmetry and topological band structure without considering net magnetic moments [20].

In ferromagnetic materials, the distinction whether the magnetization or the Berry curvature is the origin of the AHE is more subtle: While most ferromagnets show a finite AHE that is proportional to the net magnetization of the sample, the AHE in principle could be tuned to zero, irrespective of the magnetization value. For a set of bands, the Berry curvature depends on how it is connected throughout the BZ [21]. In a magnetic semiconductor without band inversion, the AHC vanishes, as the occupied valence states can be adiabatically connected to the topologically trivial vacuum (Fig. 1a). In a doped magnetic semiconductor or topologically trivial magnetic metal, the AHC can take small to large values depending on the details of the band structure; it usually follows the density of states (DOS) (Fig. 1b). The energy dependence of the AHC is constrained if there is a trivial band gap close in energy, where the AHC has to vanish. In topological metals, such as Weyl semimetals or nodal line semimetals, the connectivity of the occupied bands is nontrivial owing to the band crossings. While the Berry curvature exhibits a strong peak close to the crossing points, the DOS vanishes, and the AHC peaks sharply in energy, with giant values (Fig. 1c). The inverse relation between the DOS and AHC leads to a large anomalous Hall angle (AHA), unlike the small AHA in regular magnetic metals.

At a symmetry-protected crossing, e.g., in nodal line semimetals, the removal of the protecting symmetry leads to a gapping of the crossings, and the band structure becomes topologically equivalent to that of a semiconductor (Fig. 1d). We define a band structure as semiconducting in a topological sense, if it can be adiabatically deformed into the band structure of a gapped trivial semiconductor, even if it would be metallic from a transport perspective.

In this paper, we demonstrate these principles theoretically and experimentally and show that the AHC can be selectively tuned from 0 to 1600 $\Omega^{-1}$cm$^{-1}$ in magnetic Heusler compounds via suitable manipulations of the symmetries and band structures of the materials. In contrast to previous studies, where a finite AHE has been observed without a net magnetization, we here show that the AHE can be tuned to zero despite a finite magnetization. Magnetic Heusler compounds are excellent materials for this purpose owing to the diverse possibilities for tuning the electronic and magnetic structure by varying the composition [22,23]. The large AHC and high Curie temperature ($T_C$) allow us to obtain giant AHAs of up to 12% at room temperature.

Ternary Heusler compounds with the formula $X_2YZ$ (where $X$ and $Y$ are transition metals, and $Z$ is a main group element) crystallize in a face-centred cubic lattice with either space group (SG) $Fm\bar{3}m$ (225) (the regular Heusler structure) or SG $F\bar{4}3m$ (216) (the inverse Heusler structure). In the regular Heusler structure, the $X$ atoms occupy the Wyckoff position $8c$ $(\frac{1}{4},\frac{1}{4},\frac{1}{4})$, whereas the $Y$ and $Z$ atoms occupy the Wyckoff positions $4b$ $(\frac{1}{2},\frac{1}{2},\frac{1}{2})$ and $4a$ $(0,0,0)$ with Cu$_2$MnAl (L2$_1$) as the prototype [24]. However, in the inverse Heusler structure, atoms in the $4b$ position replace half of the $8c$ atoms, and the revised atomic arrangement becomes $X$ [$4d$ $(\frac{1}{4},\frac{1}{4},\frac{1}{4})$] $X$ [$4b$ $(\frac{1}{2},\frac{1}{2},\frac{1}{2})$] $Y$ [$4c$ $(\frac{3}{4},\frac{3}{4},\frac{3}{4})$] $Z$ [$4a$ $(0,0,0)$], with Li$_2$AgSb as the prototype. The magnetic ground state (ferromagnetic, ferrimagnetic, or antiferromagnetic) of Heusler compounds is controlled by the interatomic distances of the corresponding $X$ and $Y$ atoms and follows the Slater–Pauling rules [25,26]. Further, $T_C$ generally scales with the sum of the local moments and can reach 1200 K [27].

We compare the structures of regular and inverse Heusler compounds in Fig. 2a–2c. The L2$_1$ structure (225) of regular Heusler compounds is symmorphic, and the crystallographic point group is $O_h$. Upon changing to the inverse Heusler structure, the structure loses inversion symmetry, and the point group reduces to $T_d$. Fig. 2d–2f show the band structures along the high-symmetry points of three representative compounds. Evidently, the variation in the number of valance electrons ($N_V$) affects only the position of the Fermi energy $E_F$ within the same symmetry group of compounds, as shown for Co$_2$MnGa ($N_V$ = 28) and Co$_2$VGa ($N_V$ = 26). In both compounds, the spin-up electrons form a topological nodal line, while the spin-down electrons possess a continuous gap throughout the BZ. The Fermi level lies in the band gap for Co$_2$VGa, while a small hole pocket is formed for Co$_2$MnGa. The band structure changes only slightly as the crystal symmetry changes from 225 (Co$_2$VGa) to 216 (Mn$_2$CoGa), whereas the total magnetic moment and $N_V$ remain the same. However, this symmetry change converts the topological nodal line semimetal Co$_2$VGa into the trivial metal Mn$_2$CoGa (Fig. 1d). Being topologically nontrivial, Co$_2$VGa and Co$_2$MnGa should display an AHE with a peak close to the Fermi level (Fig. 1c). In contrast, the AHE of Mn$_2$CoGa should vanish at an energy close to the Fermi level, as it is topologically equivalent to a regular semiconductor (Fig. 1a).

Fig. 3a and 3c show the *M–H* hysteresis loops of single crystals of Co$_2$VGa and Mn$_2$CoGa, respectively, at temperatures between 2 and 300 K. The saturation magnetizations of the samples correspond to $M_S$ = 2.0 and 2.05$\mu_B$ at 2 K, respectively, and decrease with increasing temperature. Our measured $M_S$ is in good agreement with the spin magnetic moment predicted by the Slater–Pauling rule, $M_S = (N_V - 24)\mu_B$ [25,26]. Temperature-dependent magnetization measurements reveal Curie temperatures of 328 K for Co$_2$VGa and 715 K for Mn$_2$CoGa [see Fig. S4 in the supplementary information (SI)].

Having established their similar saturation magnetization, we now discuss the AHE of both compounds. The total Hall resistivity, $\rho_{yx}$ is expressed as:

$$\rho_{yx} = R_0 \mu_0 H + \rho_{yx}^{AHE}, \tag{1}$$

where $R_0$ is the linear Hall coefficient and $\rho_{yx}^{AHE}$ is the total anomalous Hall contribution. In absence of any non-collinear spin textures, $\rho_{yx}^{AHE}$ is commonly believed to scale with the spontaneous magnetization of materials [1,4,28]. However, this doesn't hold true for the topological magnetic materials. We calculate the Hall conductivity from the diagonal and off-diagonal components of the resistivity tensor as

$$\sigma_{xy} = \rho_{yx}/(\rho_{xx}^2 + \rho_{yx}^2), \qquad (2)$$

where $\rho_{xx}$ is the longitudinal resistivity. All known materials with ferromagnetic or ferrimagnetic ordering that exhibit spontaneous magnetization show an AHE [1]. Co$_2$VGa obeys this rule very well, as seen in the summary of magneto-transport measurement in Fig. 3b, with *B* along the [001] direction. $\sigma_{xy}(B)$ increases linearly for a small applied magnetic field, and anomalous behaviour is observed up to ~0.5 T. At the higher magnetic fields, the Hall resistivity slowly changes depending on $R_0$. However, at lower *B*, where the spins are not saturated, $\rho_{yx}$ is expected to have a contribution from the topological component. We estimate the AHC, $\sigma^A$, to be 137 $\Omega^{-1}$cm$^{-1}$ at 2 K by extrapolating the high field $\sigma_{xy}(B)$ data to the $B \to 0$ value. $\sigma^A$ gradually decreases at higher temperature to 58 $\Omega^{-1}$cm$^{-1}$ at 300 K. The Hall coefficient $R_0$ is calculated from the high field slope of $\rho_{yx}(B)$ as $4.6 \times 10^{-4}$ cm$^3$/C at 2 K. The sign of $R_0$ divulges important information about the type of charge carrier involved in transport. The majority charge carriers in Co$_2$VGa are clearly the hole type in the entire measured temperature range (see Fig. S5 in the SI).

In contrast, Mn$_2$CoGa shows very different anomalous Hall behaviour. Fig. 3d illustrates the Hall conductivity at various temperatures. Interestingly, $\sigma_{xy}$ increases linearly with the field, similar to a normal Hall effect. This is an exceptional observation in a metallic magnetic material with a large magnetic moment of 2 $\mu_B$ and differs markedly with other similar compounds reported in the literature [1,29]. The calculated $R_0$ is 0.035 cm$^3$/C at 2 K for Mn$_2$CoGa and decreases sharply (by a factor of 10) to 0.0035 cm$^3$/C at 300 K. Consequently, $\sigma_{xy}(B)$ decreases remarkably to a negligible value at 300 K compared to that at 2 K (Fig. 3d). The charge carriers are of the hole type. For all the above measurements, the magnetic field $B$ and current were applied along the [001] and [100] directions, respectively. However, $B$ along [011] or [$\bar{1}$11] show similar properties, indicating small anisotropy in the sample.

We now want to understand the unusual behavior of Mn$_2$CoGa theoretically. As discussed earlier, the inverse Heusler Mn$_2$CoGa can be considered as the symmetry-reduced counterpart of regular Heusler Co$_2$VGa, which has the same $N_V$. Co$_2$VGa belongs to the space group of $Fm\bar{3}m$, and the topological properties in Co$_2$VGa are due to the three mirror planes $M_x$, $M_y$ and $M_z$. Without considering spin orbit coupling, the full Hamiltonian can be decomposed into the direct product of a spin-up Hamiltonian and spin-down Hamiltonian, i.e. the band degeneracy of each spin channel is decided by the corresponding Hamiltonian. Owing to the mirror symmetry, the band inversion between the bands with opposite mirror eigenvalue forms three gapless nodal lines in the $k_x = 0$, $k_y = 0$ and $k_z = 0$ mirror planes respectively. Compared to Co$_2$VGa, the crucial difference in Mn$_2$CoGa is the absent of these three mirror planes. Though they have the same $N_V$, the mirror plane protected gapless nodal lines don't exist in Mn$_2$CoGa due to the lack of the three mirror planes.

After taking SOC into consideration, the symmetry of the system is reduced depending on the direction of magnetization, which leads to band anti-crossings of the nodal lines and generates Weyl nodes near $E_F$ for regular Heusler (like Co$_2$VGa, Co$_2$MnGa etc.) [21,30]. This is due to the fact that the mirror planes perpendicular to the magnetization do not preserve the direction of the spins, while the ones parallel do. For example, if we consider a magnetization along the $\hat{z}$ direction (as used in the experiment), the mirrors $M_x$, $M_y$ are no longer allowed symmetry operations, whereas $M_z$ remains a symmetry operation. Thus, an anti-crossing band gap will only appear from the two nodal lines in the $k_x = 0$, $k_y = 0$ planes. However, some linear crossing points are still allowed due to the combined time-reversal and rotational symmetries, which are just the Weyl points [21,30,31]. In comparison with regular Heusler compounds such as $Co_2VGa$, the absence of nodal lines in inverse Heusler compounds (like $Mn_2CoGa$) without spin–orbit interaction generally leads to the absence of Weyl points.

Fig. 4a and 4b show the computed spin-resolved DOS for $Co_2VGa$ and $Mn_2CoGa$, respectively. $Co_2VGa$ is clearly a half-metallic ferromagnet, i.e. one spin channel is gapped while the other one is metallic. In contrast, $Mn_2CoGa$ possesses one semimetallic spin channel with both electron and hole pockets. The other spin channel has a global band gap in the range of ~ $E_F$ + 0.02 to ~ $E_F$ + 0.1 eV due to the absence of the mirror symmetries. The closely related compound $Mn_2CoAl$ is a spin-gapless semiconductor, in which one spin channel is gapped, whereas the other spin channel is semimetallic with a vanishing DOS at the Fermi energy [32].

We now discuss these band structure effects on the AHC of both compounds (for details of the calculation, see the Methods section). In the limit of weak SOC, the AHC is given as the sum of the conductivities each spin species, $\sigma^A(\mu) = \sigma^{\uparrow A}(\mu) + \sigma^{\downarrow A}(\mu)$, where $\mu$ is the chemical potential [33]. For the half-metallic ferromagnet $Co_2VGa$, the states around the Fermi level arise only from the majority states, whereas the minority states exhibit a band gap of about 0.2 eV. Thus, the contribution to the AHC of the minority carriers remains constant throughout the band gap energy window (Fig. 4c). Because most of the slightly gapped nodal lines from SOC that generate a large Berry curvature, lie far from the Fermi level (Fig. 2d), the absolute value of the integrated Berry curvature of the spin-up channel is not very large. We calculate an intrinsic AHC of $\sigma^A(\mu)$

= 140 $\Omega^{-1}$cm$^{-1}$ for Co$_2$VGa, which is consistent with the experiment.

In Mn$_2$CoGa, however, both spin species possess a finite AHC around the Fermi level (Fig. 4d). The sign of the AHC of the spin-up electrons is opposite to that of the AHC of the spin-down electrons around $E_F$. Close to $E_F$, we can approximate $\sigma^A(\mu) \approx \sigma^{\uparrow A}(\mu) - |\sigma^{\downarrow A}(\mu)|$. Therefore, $\sigma^A$ can switch sign depending on which spin channel contributes the most to the overall AHC. At the charge neutrality point, the Berry curvature contribution of the spin-up channel is approximately 74 $\Omega^{-1}$cm$^{-1}$ larger than that of the spin-down channel, resulting in a finite AHC. In our ab initio calculation, the compensation point $\sigma^A(\mu) = 0$ lies about 0.1 eV above the Fermi level, which corresponds to approximately 0.1 extra electrons per formula unit. As shown in the SI, our compound is slightly electron-doped, which could explain the discrepancy between theory and experiment. More importantly, the presence of such a compensation point close to $E_F$ is a general topological property of compounds with an electronic structure topologically equivalent to that of a semiconductor, such as spin-gapless semiconductors. These occur in the inverse Heusler structure owing to breaking of the mirror symmetries if the corresponding full Heusler compound with the same $N_V$ is a half-metallic nodal line semimetal.

Since the AHC can be obtained by the integral of the Berry curvature in the Brillouin zone, we can use the Berry curvature distribution in k space of the two groups of Heusler compounds to correlate the band structure and corresponding AHC. As discussed, when Co$_2$VGa is magnetized along the z direction, the mirrors $M_x$ and $M_y$ are no longer symmetry planes, and a gapless nodal line exists only in the $k_z = 0$ plane. The band structure of a gapless nodal line contains nonzero Berry curvature around it; however, it is helically distributed in the mirror plane, and the total flux through the mirror plane is zero. Thus, the gapless nodal line in the $k_z = 0$ plane does not contribute to the intrinsic AHC of Co$_2$VGa. However, the nodal lines in the $k_x = 0$ and $k_y = 0$ planes are gapped by SOC, leading to band anti-crossings, which forces the Berry curvature to be oriented along the magnetization direction. For example, in the $k_y = 0$ plane, the $k$ points around the broken nodal lines are dominated by the negative $\Omega_{xy}$ components of the Berry curvature. The Berry curvature distribution in the Brillouin zone for $Co_2VGa$ is shown in Fig. 4e. Because the nodal lines in the mirror planes have strong energy dispersion [30,34], only some parts of the nodal lines contribute to the AHC at the fixed Fermi level, resulting in an AHC of ~140 $\Omega^{-1}cm^{-1}$ at $E_F$.

The Berry curvature distribution in $Mn_2CoGa$ is very different due to the absence of a nodal line band structure (Fig. 4f). In $Mn_2CoGa$ it contains both positive and negative hot spots, throughout the Brillouin zone. Therefore, during integration of the Berry curvature, hot spots of opposite sign cancel, producing a small AHC value. By doping or temperature effects, one can tune the chemical potential to the topologically required critical point where the positive and negative Berry curvatures are totally cancelled out, yielding *zero* AHC. Interestingly, this is validated by our experiment, as shown in the inset of Fig. 3d for $Mn_2CoGa$. As the temperature increases up to 300 K, a nonzero AHE is observed, and $\rho_{yx}(B)$ clearly scales with the sample's magnetization.

The obtained results show that the interplay of the crystal symmetry with the topological and geometrical properties of the Berry curvature provides a powerful framework to control the AHC, independent of the magnetization [17,33]. We condense our findings in Fig. 5, which presents a strategy for tuning the electronic and topological properties of Heusler compounds. In regular Heusler compounds, we find a half-metallic nodal line protected by mirror symmetry at different valence electron counts. By suitable chemical substitution, one may destroy the mirror symmetry of the compound (Fig. 5a). The removal of certain mirror symmetries changes the electronic structure of a half-metallic nodal line semimetal into a trivial band structure very close to that of a spin-gapless semiconductor by removing the nodal lines. These changes induce corresponding changes in the Berry curvature and thus lead to a small or vanishing AHC. Because of the excellent tunability, one can easily manipulate the chemical potential in Heusler compounds by changing $N_V$ (Fig. 5b). For example, $\sigma^A$ changes from ~140 $\Omega^{-1}$cm$^{-1}$ for Co$_2$VGa ($N_V = 26$) to ~1600 $\Omega^{-1}$cm$^{-1}$ for Co$_2$MnGa ($N_V = 28$), which also possesses nodal lines around the Fermi level, although the compounds have different $N_V$.

Again, keeping the same $N_V$, when the crystal symmetry is altered from $O_h$ in Co$_2$VGa to $T_d$ in Mn$_2$CoGa, $\sigma^A$ decreases to zero (Fig. 5c). The maximum AHA, $\Theta_{AHE} = \Delta\sigma^A/\sigma_{xx}$, reaches a giant value of ~12% for Co$_2$MnGa at room temperature, at which the nodal line is closest to $E_F$ (Fig. 5d). We conjecture that the nodal line dispersion and charge carrier concentration (Fig. 2d and 2e) control the value of $\Theta_{AHE}$, which is why $\Theta_{AHE}$ for Co$_2$VGa is only about 2% (Fig. 5d).

Owing to the extensive tunability of Heusler compounds and the topological nature of our arguments, the compensated AHC should not be unique to Mn$_2$CoGa. Since all the spin-gapless Heusler semiconductors have similar Berry curvature distributions, the same principle generally applies among all of them, for example, Ti$_2$CrSi, CoVScSi, and CoFeMnSi. In the SI, we list a series of spin-gapless semiconductor candidate compounds, classified according to the corresponding $N_V$ that should show a compensated AHC near $E_F$. Therefore, we can add a new rule to the simple rules of Heusler compounds [24]: For each spin-gapless semiconductor, there exists a corresponding half-metallic full Heusler compound with a nodal line in the majority spin channel and Weyl nodes close to $E_F$ with the same $N_V$, and vice versa. This implies that there should be many more compounds with a large AHC and AHA in the Heusler family. The AHC of the full Heusler compound will generally be large, whereas that of the inverse Heusler compound will be close to zero, as we illustrate for CoFeMnSi and Co$_2$MnGa, which have $N_V = 28$ (Fig. 5g and 5h, respectively).

Vice versa to previous studies, where a finite AHE has been observed without a net magnetization, we have showed that the AHE can be tuned to zero despite a finite magnetization. Thus, our results complement the recent investigations on the Berry curvature origin of the AHE. Our findings are general and can be extended to other classes of materials with finite magnetization, as one can control the topology by changing the magnetic space group. Thus, our work is relevant to the recent interest in topological classification of magnetic materials, as the topology often constrains the Berry curvature distribution [6]. By symmetry engineering, a metallic magnet can be converted to a topologically trivial semiconductor with zero AHC. The possibility to tune the AHE from *zero* to a colossal value, independent of the magnetization of the material, may be interesting for next-generation topo-spintronics applications. Additionally, topological semimetals with a high Curie temperature and large AHA, such as Heusler compounds, are excellent candidate materials for a confinement-induced quantum AHE in thin films.

**Methods**

Single crystals of $Mn_2CoGa$, $Co_2VGa$, and $Co_2MnGa$ were grown using the Bridgman–Stockbarger crystal growth technique. First, stoichiometric amounts of high-purity metals were premelted in an alumina crucible using induction melting. Then the crushed powder was packed in a custom-designed sharp-edged alumina tube, which was sealed in a tantalum tube. Before crystal growth, the compound's melting point was determined using differential scanning calorimetry measurement. The samples were heated (to 1150 °C for $Mn_2CoGa$, 1450 °C for $Co_2VGa$, and 1250 °C for $Co_2MnGa$), held there for 10 h to ensure homogeneous mixing of the melt, and then slowly cooled to 750 °C. Single crystallinity was checked by white-beam backscattering Laue X-ray diffraction at room temperature. The crystal structures were analysed with a Bruker D8 VENTURE X-ray diffractometer using Mo-K radiation. Magnetization measurements were performed using a Quantum Design vibrating sample magnetometer (MPMS). The transport properties were characterized by a Quantum Design physical property measurement system (PPMS, ACT option).

The electronic band structure was calculated using density functional theory (DFT) with the localized atomic orbital basis and the full potential as implemented in the full-potential local-orbital (FPLO) code [35]. The exchange and correlation energies were considered in the generalized gradient approximation (GGA), following the Perdew-Burke-Ernzerhof parametrization scheme [36]. To calculate the AHC, we projected the Bloch wave functions into high-symmetry atomic-like Wannier functions, and constructed the tight-binding model Hamiltonian. On the basis of the tight-binding model Hamiltonian, we calculated the AHC using the Kubo formula and clean limit [14]:

$$\sigma_{xy}^z(E_F) = \frac{e^2}{\hbar}(\frac{1}{2\pi})^3 \int_{\vec{k}} d\vec{k} \sum_{E(n,\vec{k})<E_F} f(n,\vec{k}) \Omega_{n,xy}^z(\vec{k})$$

$$\Omega_{n,xy}^z(\vec{k}) = \text{Im} \sum_{n' \neq n} \frac{<u(n,\vec{k})|\hat{v}_x|u(n',\vec{k})><u(n',\vec{k})|\hat{v}_y|u(n,\vec{k})> - (x \leftrightarrow y)}{(E(n,\vec{k}) - E(n',\vec{k}))^2}$$

where $f(n,\vec{k})$ is the Fermi-Dirac distribution, $E(n,\vec{k})$ is the eigenvalue for the *n*-th eigen states of $|u(n,\vec{k})>$ at the $\vec{k}$ point, and $\hat{v}_{x(y)}$ is the velocity operator. A $201\times 201\times 201$ *k*-grid was used in the integral. We also classified the Berry curvature $\Omega_{n,xy}^z(\vec{k})$ into two species, $\Omega_{n,xy}^{z,\uparrow}(\vec{k})$ and $\Omega_{n,xy}^{z,\downarrow}(\vec{k})$, using the expectation value $\hat{s}_{n,z}(\vec{k})$ of the wavefunction for the given band and *k* point. The spin-resolved Berry phases $\sigma_{xy}^{z,\uparrow}(E_F)$ and $\sigma_{xy}^{z,\downarrow}(E_F)$ are obtained by taking the integral of $\Omega_{n,xy}^{z,\uparrow}(\vec{k})$ and $\Omega_{n,xy}^{z,\downarrow}(\vec{k})$, respectively, in the entire BZ.

**Supplementary Information** is linked to the online version of the paper.

**Acknowledgements**

This work was financially supported by the ERC Advanced Grant No. 291472 `Idea Heusler', ERC Advanced Grant No 742068 – TOPMAT, and Deutsche Forschungsgemeinschaft DFG under SFB 1143. T.H.K. acknowledges financial support from the Ministry of Science and Technology, Taiwan, under Grant No. 105-2917-I-110-004 and 103-2112-M-110 -010 -MY3. L.M. is grateful for the hospitality of the MPI CPFS, where part of the work was conducted.

**Competing financial interests**

The authors declare no competing financial interests.

**Correspondence**

\* Correspondence and requests for materials should be addressed to K. Manna (email: kaustuvmanna@gmail.com) & C. Felser (email: felser@cpfs.mpg.de).

[†] Present address: Center for Computational Quantum Physics, The Flatiron Institute-New York, New York, 10010, USA.

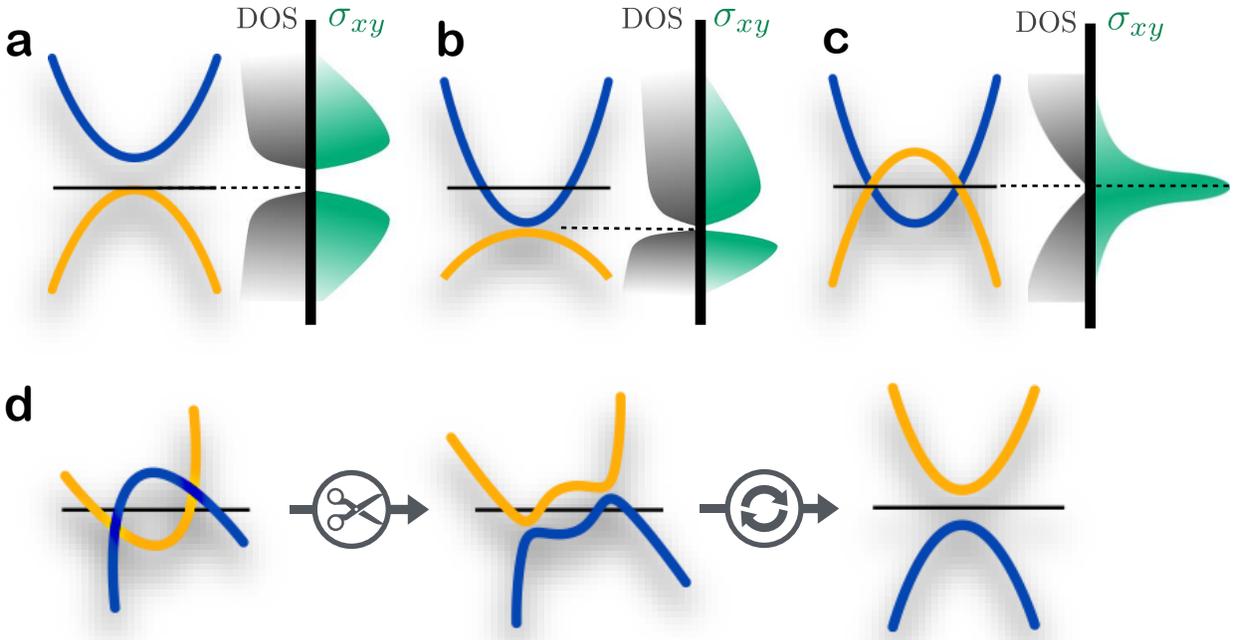

Fig. 1: Band structure, density of states and AHC of (a) a trivial magnetic semiconductor, (b) a trivial metal, and (c) a topological metal. (d) Change in the topology of a topological metal by gapping out the crossing points via symmetry-breaking or other strong perturbations. The gapped band structure is topologically equivalent to that of a trivial semiconductor.

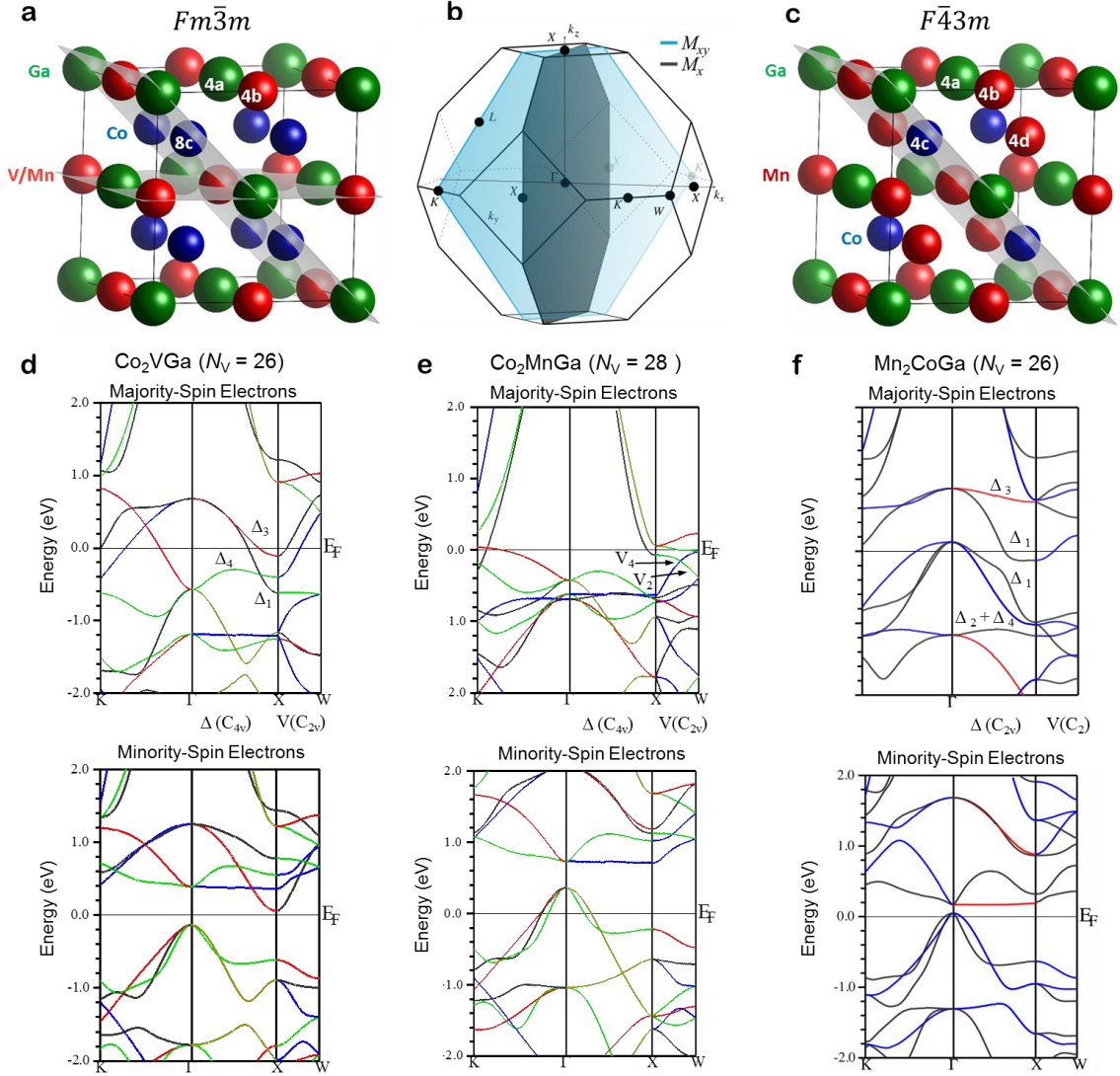

Fig. 2: (a) Lattice arrangements and (b) first bulk BZ for the regular Heusler compounds (Co$_2$VGa and Co$_2$MnGa) with SG $Fm\bar{3}m$. Corresponding high-symmetry points of the BZ are indicated, along with the allowed mirror symmetries. (c) Lattice arrangements for the inverse Heusler compounds (Mn$_2$CoGa) with SG $F\bar{4}3m$. The only allowed type of mirror plane is highlighted. Computed band structure of (d) Co$_2$VGa, (e) Co$_2$MnGa, and (f) Mn$_2$CoGa. Colours represent different irreducible representations (denoted by Δ and V) of the little groups at each *k*.

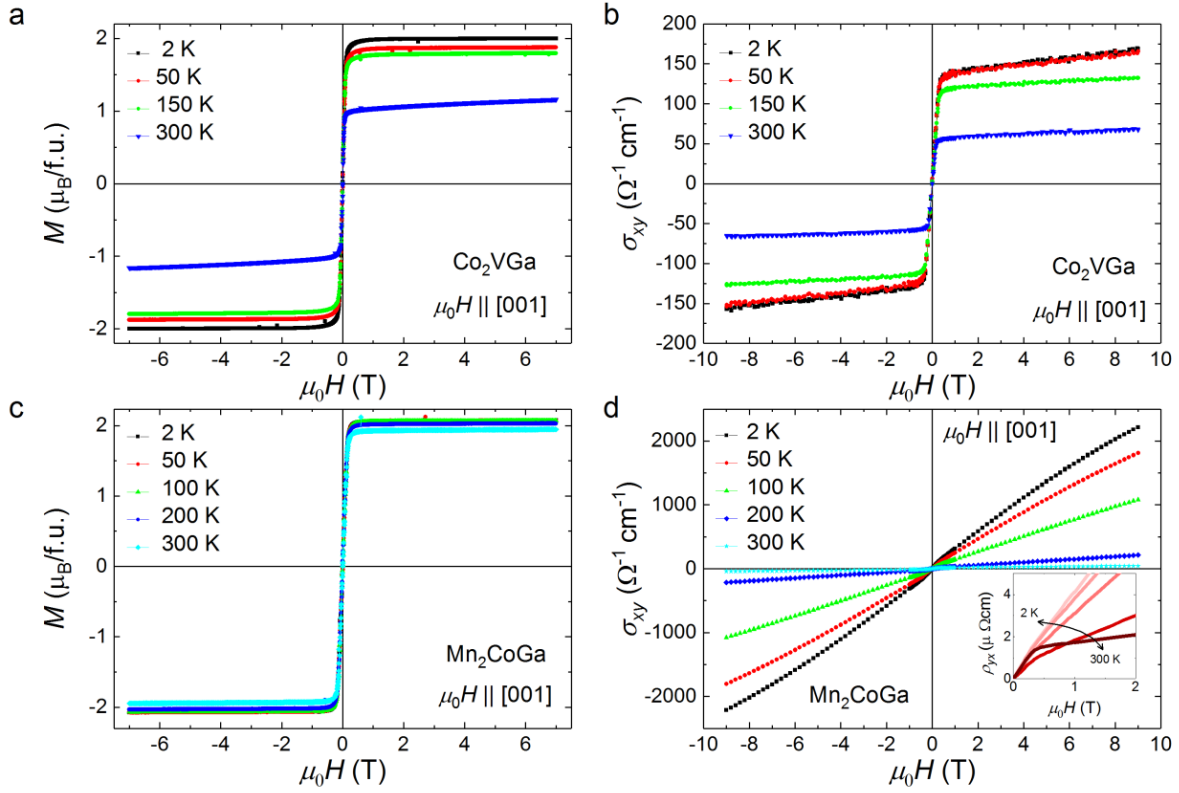

Fig. 3: (a), (c) Magnetic-field-dependent magnetization and (b), (d) corresponding field- and temperature-dependent AHC [$\sigma_{xy}$] of (a), (b) $Co_2VGa$ and (c), (d) $Mn_2CoGa$ single crystals with field along [001] direction. Inset of (d) shows a magnified view of the field-dependent Hall resistivity [$\rho_{yx}$] of $Mn_2CoGa$ at various temperatures from 2 to 300 K in the low-field region.

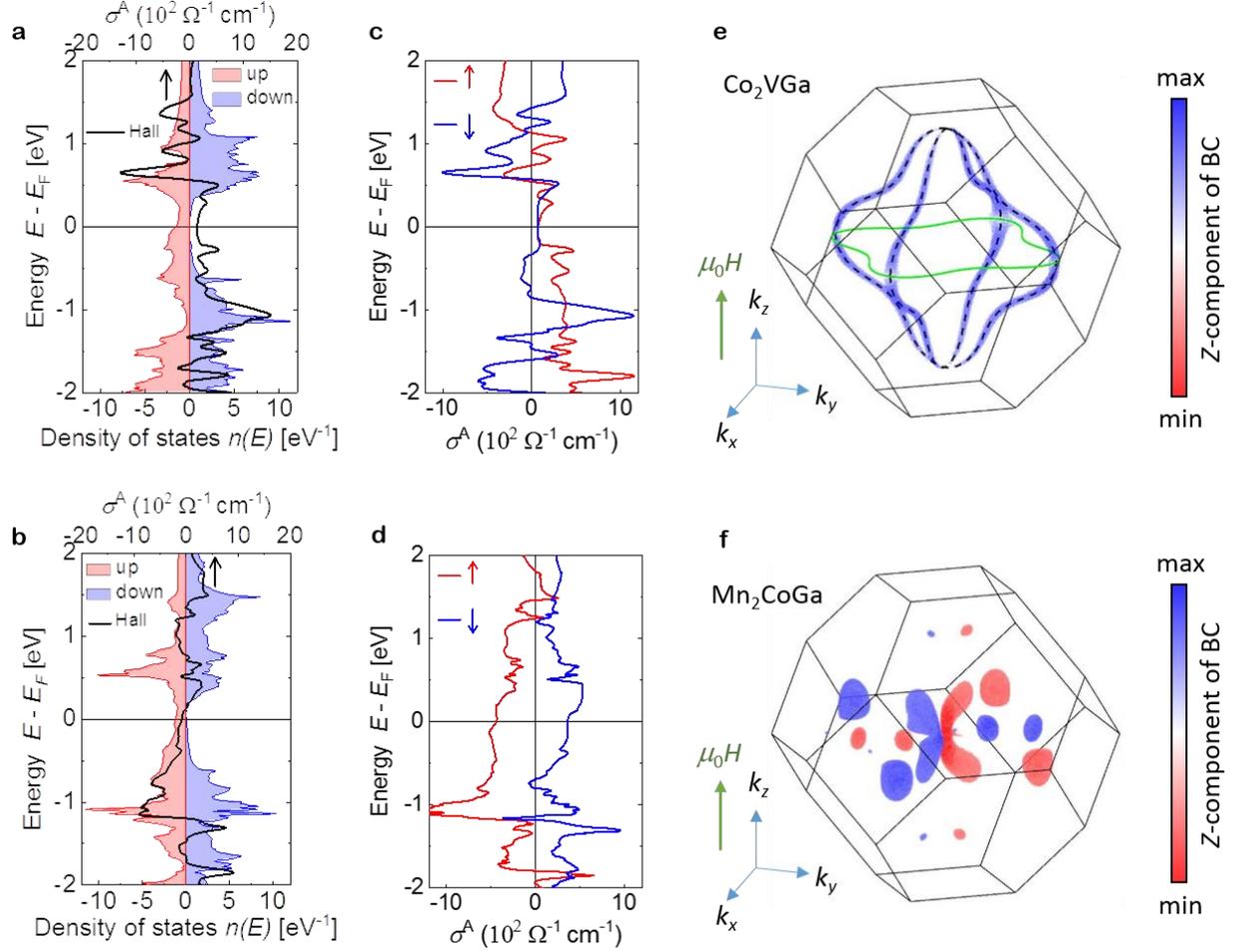

Fig. 4: Computed spin-resolved DOS for (a) $Co_2VGa$ and (b) $Mn_2CoGa$. Black lines represent the calculated net anomalous Hall conductivity ($\sigma^A$) values. Red and blue represent the DOS for the majority and minority carriers, respectively. Hall contributions in the majority and minority spin channels for (c) $Co_2VGa$ and (d) $Mn_2CoGa$. Berry curvature distribution in the BZ for (e) $Co_2VGa$ and (f) $Mn_2CoGa$ calculated for all valence bands. The solid green line in (e) represents the protected nodal line due to the $M_z$ symmetry operation with magnetic field along $z$-direction. The dotted black lines are the gapped nodal lines which the Berry curvature (BC) distribution in $Co_2VGa$ follows.

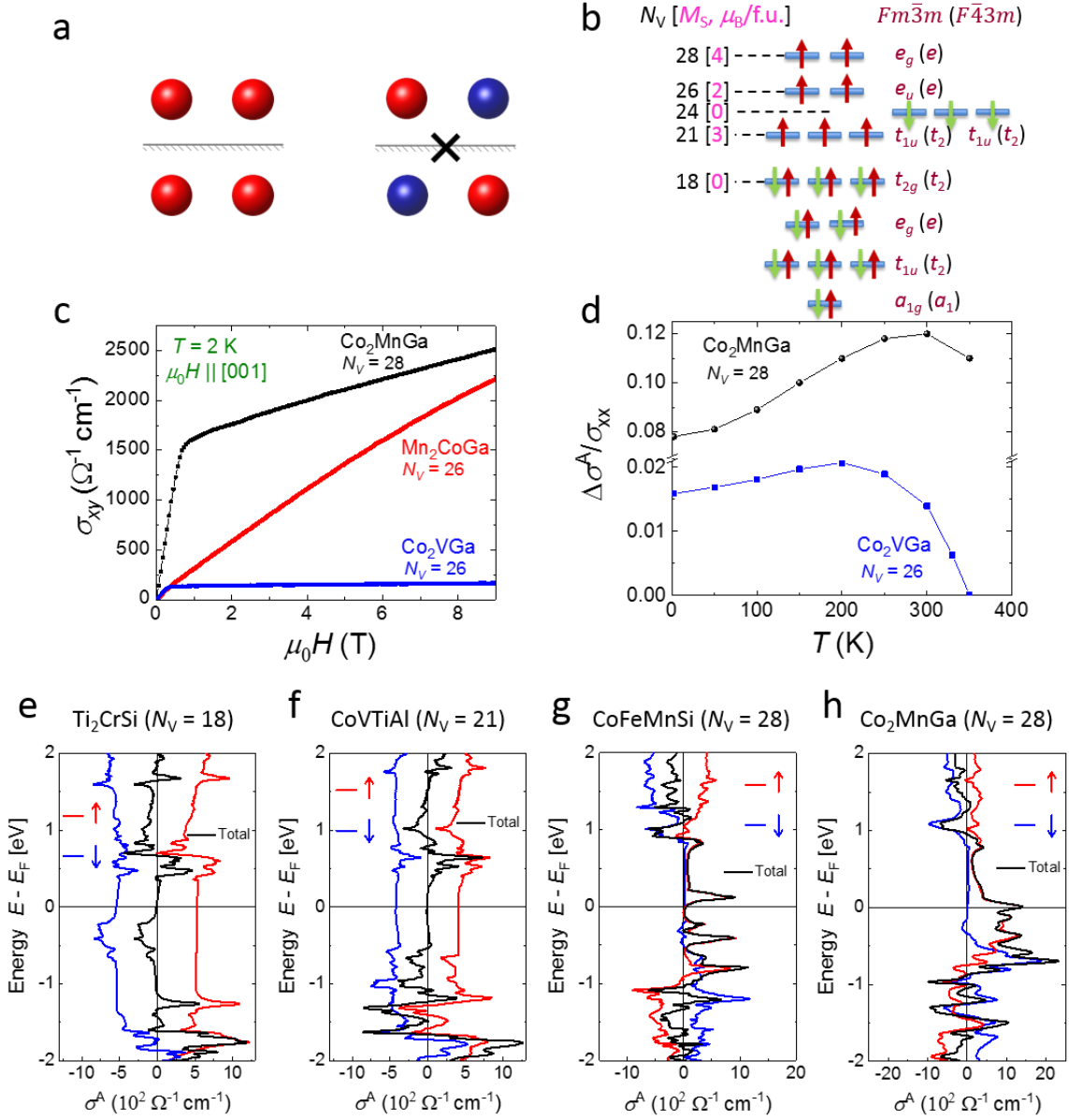

Fig. 5: (a) Breaking of mirror symmetry by chemical substitution in the compound. (b) Schematic representation of the distribution of valence electrons in various electronic energy levels for both spin channels and the corresponding net magnetic moment. (c) Experimentally observed Hall conductivity of three magnetic Heusler compounds with various $N_V$. (d) Temperature dependence of the anomalous Hall angle for two regular Heusler compounds, Co$_2$MnGa and Co$_2$VGa. Hall contributions of the majority and minority spin channels and the net anomalous Hall conductivity for (e) Ti$_2$CrSi, (f) CoVTiAl, (g) CoFeMnSi, and (h) Co$_2$MnGa.

# From colossal to zero: Controlling the Anomalous Hall Effect in Magnetic Heusler Compounds

Kaustuv Manna[1*], Lukas Muechler[1,2,†], Ting-Hui Kao[1,3], Rolf Stinshoff[1], Yang Zhang[1], Johannes Gooth[1], Nitesh Kumar[1], Guido Kreiner[1], Klaus Koepernik[4,] Roberto Car[2], Jürgen Kübler[1,5], Gerhard H. Fecher[1], Chandra Shekhar[1], Yan Sun[1], Claudia Felser[1,*]

[1]*Max Planck Institute for Chemical Physics of Solids, 01187 Dresden, Germany.*

[2]*Department of Chemistry, Princeton University, Princeton, New Jersey 08544, USA.*

[3]*Department of Physics, National Sun Yat-Sen University, Kaohsiung 804, Taiwan.*

[4]*IFW Dresden, P.O. Box 270116, 01171 Dresden, Germany.*

[5]*Institut für Festkörperphysik, Technische Universität Darmstadt, D-64289 Darmstadt, Germany.*

**Metallographic Characterization:**

The metallographic analysis was performed using an optical (Fig. S1) and a scanning probe electron (Fig. S2) microscopy for all the single crystals of $Co_2VGa$ (a), $Co_2MnGa$ (b) and $Mn_2CoGa$ (c). The samples are found to be single phase with homogeneous distribution of the chemical composition. The results of the compositional analysis with energy-dispersive X-ray (EDX) spectroscopy is summarized in table-S1. For the optical microscopy measurement, small crystals were embedded in epoxy resin blocks, and a smooth surface was prepared. In order to resolve the discrepancy between the theoretical calculation and the experimental Hall conductivity of the inverse Heusler compound $Mn_2CoGa$, a detailed chemical analysis was performed. The chemical composition of the grown $Mn_2CoGa$ single crystals was determined as: $Mn_{2.07}Co_{1.04}Ga$ confirming slight electron doping in the samples.

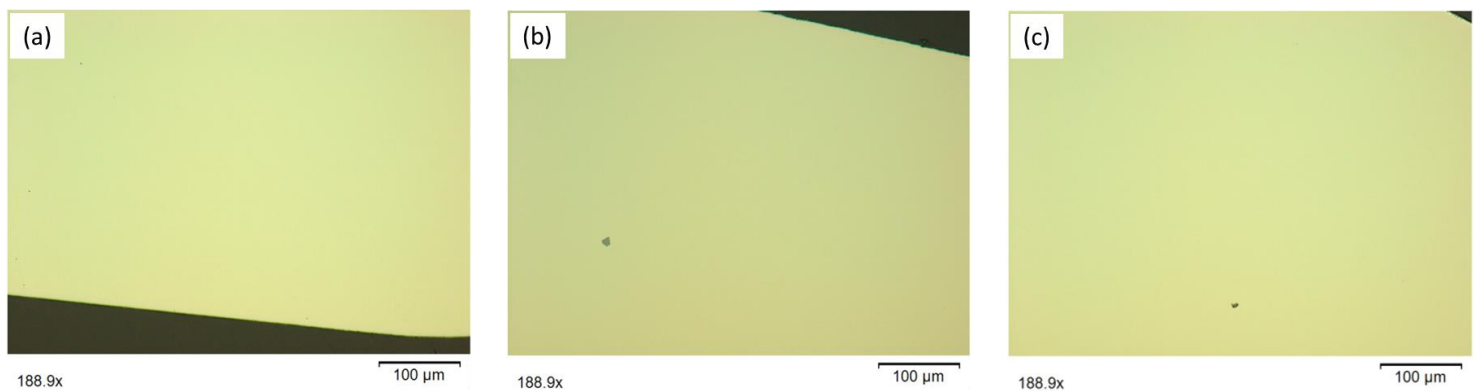

**Fig. S1: Metallographic analysis to investigate the chemical homogeneity.** Optical microscopy image for the homogeneous phase of (a) $Co_2VGa$, (b) $Co_2MnGa$ and (c) $Mn_2CoGa$ single crystals.

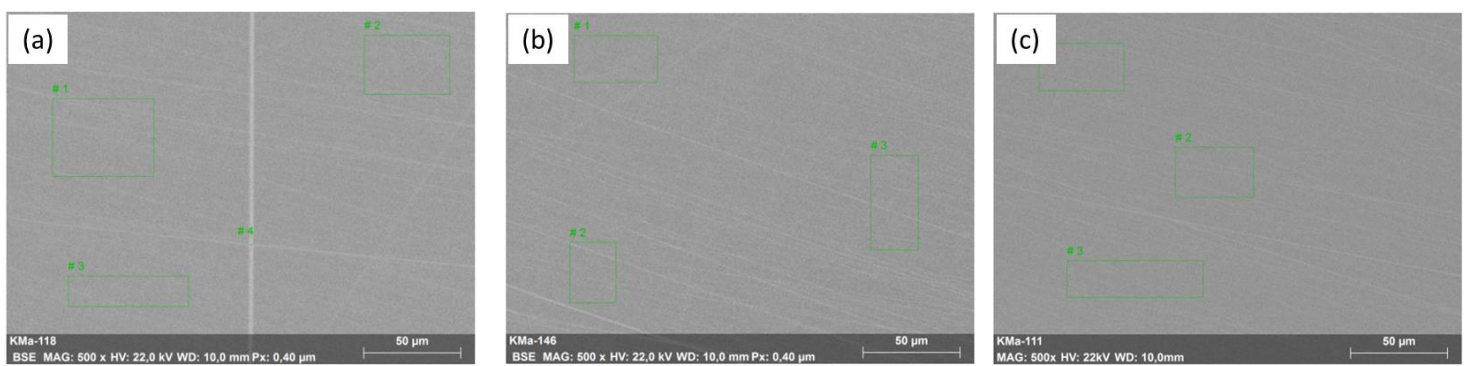

**Fig. S2: Compositional analysis with energy-dispersive X-ray (EDX) spectroscopy.** Scanning electron microscopy images for the (a) $Co_2VGa$, (b) $Co_2MnGa$ and (c) $Mn_2CoGa$ single crystals. The compositional analysis performed at the marked places and the results are summarized in table-S1.

Table-S1: Results of EDX analysis for the $Co_2VGa$, $Co_2MnGa$ and $Mn_2CoGa$ single crystals performed at the marked places in Fig. S2.

| Spot | $Co_2VGa$ (at. %) | | | $Co_2MnGa$ (at. %) | | | $Mn_2CoGa$ (at. %) | | |
|---|---|---|---|---|---|---|---|---|---|
| | Co | V | Ga | Co | Mn | Ga | Co | Mn | Ga |
| (1) | 53.27 | 23.65 | 23.07 | 52.16 | 23.50 | 24.34 | 25.84 | 49.89 | 24.28 |
| (2) | 53.12 | 23.78 | 23.10 | 52.19 | 23.53 | 24.28 | 25.96 | 49.65 | 24.39 |
| (3) | 53.22 | 23.68 | 23.10 | 52.23 | 23.61 | 24.17 | 25.95 | 49.80 | 24.25 |
| (4) | 53.14 | 23.92 | 22.94 | | | | | | |

**Crystal Orientation:**

The quality of the grown Mn$_2$CoGa, Co$_2$VGa and Co$_2$MnGa single crystals was analyzed at room temperature by a white beam backscattering Laue X-ray diffractometer and a Bruker D8 VENTURE x-ray diffractometer using Mo-K radiation. The crystals were cut in a plate-like shape and fixed with glue on a glass slide for the measurement. All samples show sharp and well-defined Laue spots that can be indexed with a single pattern, suggesting excellent quality of the grown crystals. Fig. S3 presents the measured Laue diffraction patterns of the crystals superposed with simulated ones.

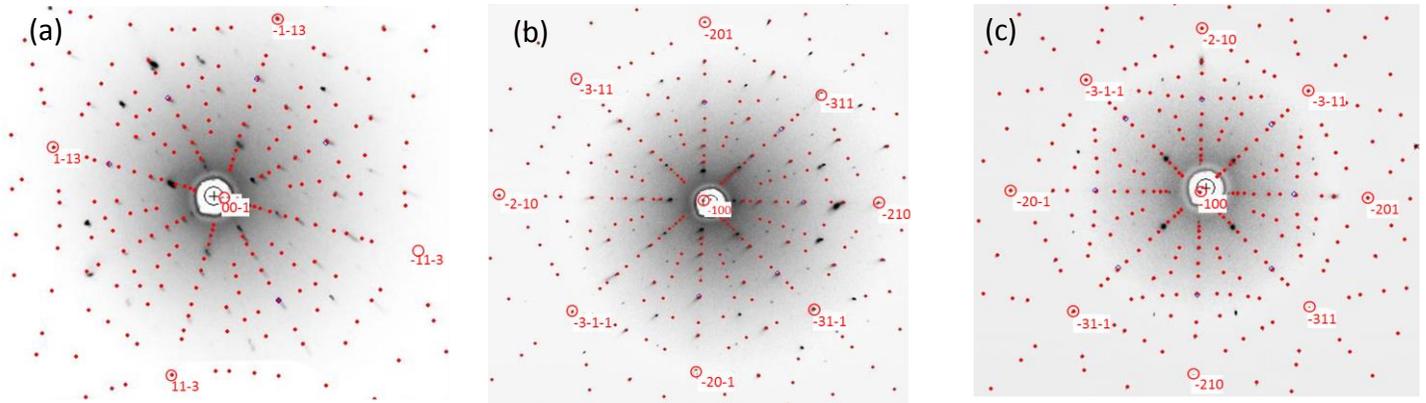

Fig. S3: Laue diffraction patterns of the oriented single crystals of (a) Co$_2$VGa, (b) Co$_2$MnGa and (c) Mn$_2$CoGa. The black spots are the observed pattern and the red spots are the calculated pattern along with Miller indices.

**Magnetization:**

The magnetization measurements were performed in a commercial Quantum Design, vibrating sample magnetometer. Fig. S4 (a), (b) and (c) presents the temperature dependent field-cooled (FC) and zero-field-cooled (ZFC) magnetization for Co$_2$VGa, Co$_2$MnGa and Mn$_2$CoGa single crystals with an applied *dc* field of $H_{dc} = 200$ Oe along the [001] direction. The magnetization sharply decreases at the Curie temperature ($T_C$) as result of the ferromagnetic to paramagnetic phase transition. The $T_C$'s of the samples were determined as 327.7 K for $Co_2VGa$, 686.4 K for $Co_2MnGa$ and 715.5 K for $Mn_2CoGa$, respectively.

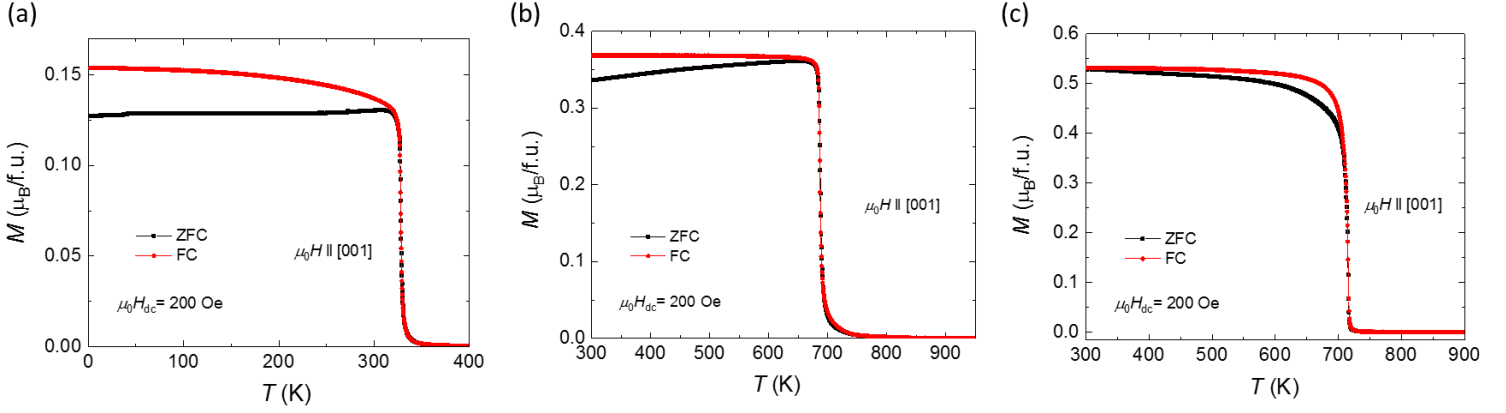

Fig. S4: Temperature dependent FC-ZFC magnetization at 200 Oe applied *dc* field for (a) $Co_2VGa$, (b) $Co_2MnGa$ and (c) $Mn_2CoGa$ single crystals.

**Electrical resistivity:**

The electrical resistivity measurements were performed in a commercial Quantum Design, physical property measurement system equipped with the AC transport option. The crystals were cut in a rectangular bar shape in desired crystallographic directions. For resistivity measurements four-probe contacts geometry and for Hall resistivity, five-probe contacts geometry were made by 25 μm Pt wire and spot-welded. Fig. S5 (a), (b) and (c) presents the temperature dependent longitudinal resistivity $\rho_{xx}(T)$ for $Co_2VGa$, $Co_2MnGa$ and $Mn_2CoGa$ respectively. Evidently, all the samples show metallic behavior with residual resistivity, $\rho_0(2\ K)$ as $1.05 \times 10^{-4}$ Ω cm for $Co_2VGa$, $5.9 \times 10^{-5}$ Ω cm for $Co_2MnGa$ and $8.8 \times 10^{-5}$ Ω cm for $Mn_2CoGa$. The residual resistivity ratio ($\rho_{300\ K}/\rho_{2\ K}$) is determined as 1.99 for $Co_2VGa$, 2.15 for $Co_2MnGa$ and 2.88 for $Mn_2CoGa$ single crystals. The resistivity of $Co_2VGa$ exhibits a typical bend at the Curie temperature.

Fig. S5 (d), (e) and (f) presents the field dependent Hall resistivity at $\rho_{xy}(B)$ at various temperatures for the Co$_2$VGa, Co$_2$MnGa and Mn$_2$CoGa single crystals respectively. The Hall coefficient, $R_0$ is calculated from the high field slope of $\rho_{yx}(B)$. All the samples possess hole type charge carriers in the entire measured temperature range. The carrier concentration is calculated using, $n = 1/(e.R_0)$ and the mobility using, $\mu_{el} = R_0/\rho_{xx}$, where $e$ is the electron charge. At 2 K, we estimate the carrier concentration as $1.36 \times 10^{22}$ cm$^{-3}$ for Co$_2$VGa, $2.3 \times 10^{21}$ cm$^{-3}$ for Co$_2$MnGa and $1.8 \times 10^{20}$ cm$^{-3}$ for Mn$_2$CoGa. The corresponding mobility at 2 K is determined as 4.36 cm$^2$ V$^{-1}$ s$^{-1}$ for Co$_2$VGa, 54.9 cm$^2$ V$^{-1}$ s$^{-1}$ for Co$_2$MnGa and 301.7 cm$^2$ V$^{-1}$ s$^{-1}$ for Mn$_2$CoGa.

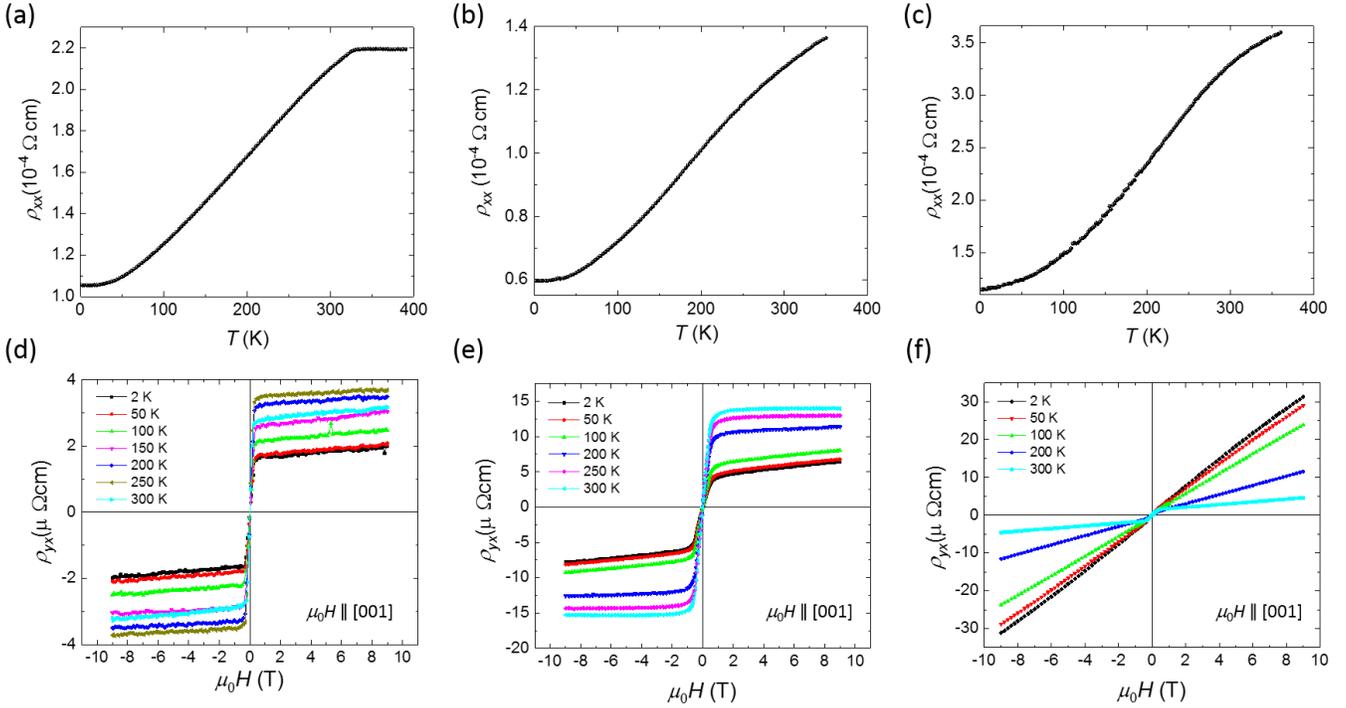

Fig. S5: Temperature dependent longitudinal resistivity for (a) Co$_2$VGa, (b) Co$_2$MnGa and (c) Mn$_2$CoGa single crystals at few selected temperatures. Field dependent Hall resistivity $\rho_{yx}(B)$ at various temperatures of (d) Co$_2$VGa, (e) Co$_2$MnGa and (f) Mn$_2$CoGa.

Field dependence of the longitudinal resistivity of all the samples were measured in the temperature window from 2 to 300 K with magnetic field upto 9 T. In Fig. S6 we compare the magnetoresistance (MR) of all the samples with *I* along [100] and *B* along [001]. The regular - Heusler ferromagnetic compounds $Co_2VGa$ and $Co_2MnGa$ shows very low negative MR as usually observed in other typical ferromagnetic systems. In contrast, $Mn_2CoGa$ shows a remarkable behavior. At lower temperature, we observe a positive MR and with increasing temperature, the MR changes sign above ~ 200 K. At lower temperatures, the MR shows a non-saturating nearly linear behavior with maximum positive MR of ~ 12.9 % at 2 K upto 9 T magnetic field. This quantum linear MR is the characteristic signature of the spin gapless semiconductor compounds as reported earlier [1,2].

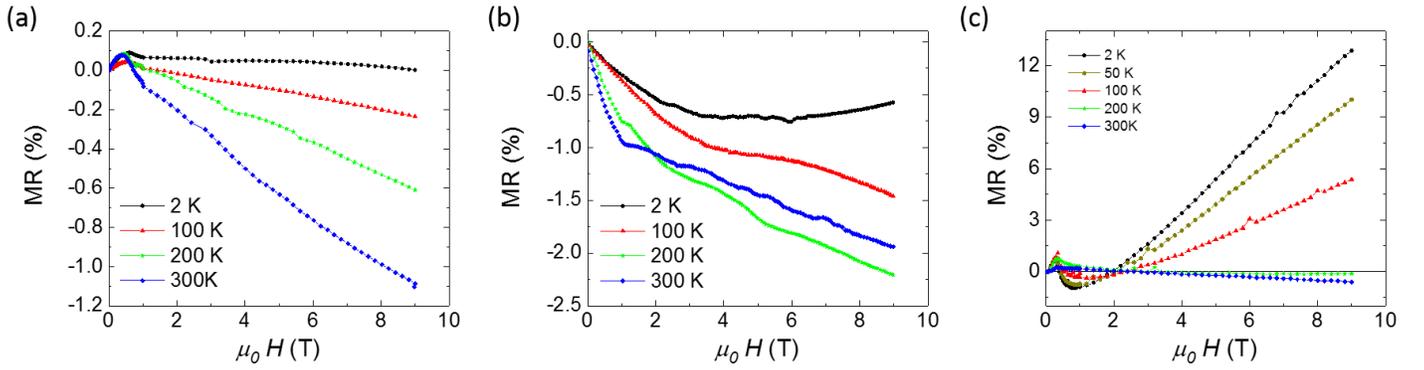

Fig. S6: Magnetic field dependence of the longitudinal magnetoresistance (MR) estimated as: $[\rho_{yx}(B) - \rho_{yx}(0)]/\rho_{yx}(0)$ at few selected temperatures for (a) $Co_2VGa$, (b) $Co_2MnGa$ and (c) $Mn_2CoGa$ single crystals.

In order to estimate the intrinsic anomalous Hall conductivity ($\sigma_{xy}^{int}$) contribution in the regular-Heusler compounds $Co_2VGa$ and $Co_2MnGa$, which is free from the extrinsic scattering effects, we fit the experimental anomalous hall resistivity ($\rho^A$) data with the longitudinal Hall resistivity ($\rho_{xx}$) using the equation:

$$\rho^A = f(\rho_{xx0}) + \sigma_{xy}^{int}\rho_{xx}^2. \qquad (1)$$

Here $f(\rho_{xx0})$ is a function of the residual resistivity $\rho_{xx0}$, which incorporates the contributions due to the skew scattering as-well-as the side jump effects, and $\sigma_{xy}^{int}$ accounts for the anomalous Hall signal purely due to the Berry curvature effect [3,4]. Here $\rho^A$ at each temperature is estimated by extrapolating the high field data of $\rho_{xx0}(B)$ to the $B \to 0$ limit. Now $T_C$ of $Co_2VGa$ is ~ 327 K and from its $\rho_{yx}(B)$ curve in fig. S5(d) we see that the $\rho^A$ increases up to ~ 250 K as normally observed in other compounds. However, above ~ 250 K we believe that the magnetic transition comes into play and $\rho^A$ decreases at higher temperature. So for a good comparison, the scaling plot between $\rho^A$ vs $\rho_{xx}^2$ in Fig. S7 is restricted in the temperature window 2 to 250 K. Evidently the experimental data fits very well in this temperature window with the linear relation of equation (1). From the slope we estimated $\sigma_{xy}^{int}$ as 1164 $\Omega^{-1}cm^{-1}$ for $Co_2MnGa$ and 95 $\Omega^{-1}cm^{-1}$ for $Co_2VGa$ single crystals.

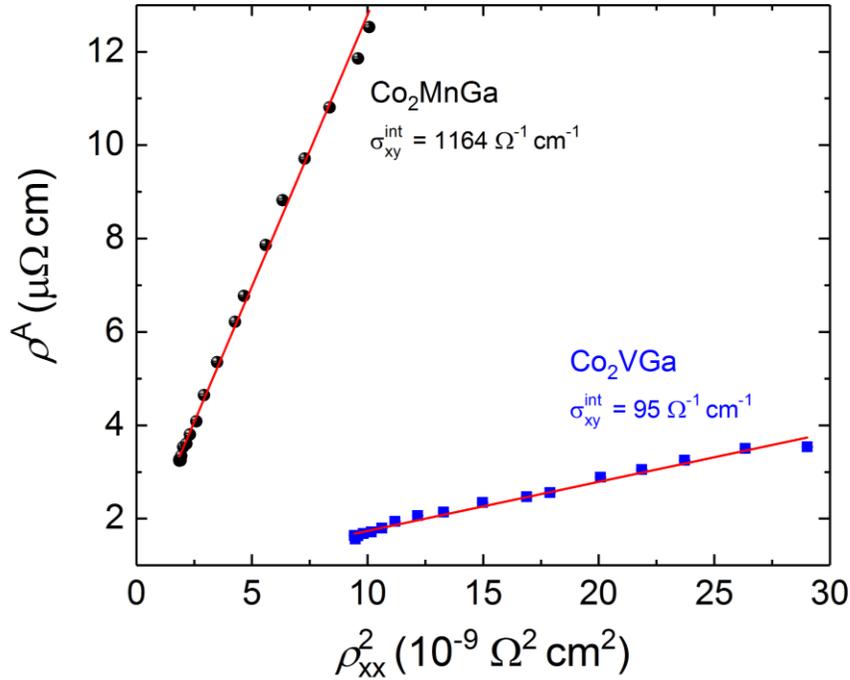

Fig. S7: Scaling relation between the anomalous Hall resistivity $\rho^A$ and $\rho_{xx}^2$ for the $Co_2MnGa$ and $Co_2VGa$ single crystals. The corresponding slope divulges the intrinsic AHC, $\sigma_{xy}^{int}$.

**More compounds:**

Here we list a series of suggested spin-gapless semiconducting compounds in Heusler family in Table-S2 with various number of valance electrons ($N_V$) that should show the compensated anomalous Hall conductivity near $E_F$.

Table-S2

| $N_V = 18$ | $N_V = 21$ | $N_V = 26$ | $N_V = 28$ |
|---|---|---|---|
| Ti$_2$CrSi | CoVTiAl | Mn$_2$CoAl | CoFeMnSi |
| Ti$_2$VAs | CoVScSi | CoFeCrAl | Cr$_2$ZnGe |
| Zr$_2$MnAl | FeCrScSi | CoMnCrSi | Cr$_2$ZnSi |
|  | FeVTiSi | CoFeVSi | Cr$_2$ZnSn |
|  | FeMnScAl | FeMnCrSb |  |
|  | FeVTiSi | CoFeCrGa |  |